\documentclass[aps,twocolumn,superscriptaddress,floatfix,prl]{revtex4-1}
\setcitestyle{super}

\usepackage{lipsum}
\usepackage{graphicx}
\usepackage{amsmath,amssymb}
\usepackage{braket}
\usepackage{mathtools}
\usepackage{hyperref}
\usepackage{multibib}

\usepackage{color}
\usepackage[normalem]{ulem}
\usepackage{amsfonts}

\usepackage{pdfpages} 
\makeatletter
\AtBeginDocument{\let\LS@rot\@undefined}
\makeatother

\definecolor{mypink1}{rgb}{0.858, 0.188, 0.478}
\definecolor{dartmouthgreen}{rgb}{0.05, 0.5, 0.06}
\hypersetup{
    colorlinks=true,
    linkcolor=blue,
    filecolor=blue,      
    urlcolor=mypink1,
    citecolor=dartmouthgreen,
}

\usepackage{pdfpages} 
\makeatletter
\AtBeginDocument{\let\LS@rot\@undefined}
\makeatother

\begin{document}
\title{Mean-field entanglement transitions in random tree tensor networks}

\author{Javier Lopez-Piqueres, Brayden Ware and Romain Vasseur}
\affiliation{Department of Physics, University of Massachusetts, Amherst, Massachusetts 01003, USA }

\date{\today}
\begin{abstract}

Entanglement phase transitions in quantum chaotic systems subject to projective measurements and in random tensor networks have emerged as a new class of critical points separating phases with different entanglement scaling. We propose a mean-field theory of such transitions by studying the entanglement properties of random tree tensor networks. As a function of bond dimension, we find a phase transition separating area-law from logarithmic scaling of the entanglement entropy. Using a mapping onto a replica statistical mechanics model defined on a Cayley tree and the cavity method, we analyze the scaling properties of such transitions. Our approach provides a tractable, mean-field-like example of an entanglement transition. We verify our predictions numerically by computing directly the entanglement of random tree tensor network states.

\end{abstract}

\maketitle
\noindent

Quantum entanglement has become an invaluable tool for studying the equilibrium and non-equilibrium properties of many-body quantum systems~\cite{RevModPhys.80.517,LAFLORENCIE20161}. Recently, a new class of phase transitions separating phases with dramatically different entanglement features has been discovered. An example of such an {\em entanglement transition} that has attracted a lot of attention is the many-body localization transition, where many-body eigenstates change from volume-law to area-law entanglement scaling as disorder is increased~\cite{doi:10.1146/annurev-conmatphys-031214-014726, Vasseur_2016, ALET2018498, RevModPhys.91.021001}. A fundamentally different class of models also exhibiting entanglement transitions is realized by chaotic quantum systems subjected to random local projective measurements~\cite{2019ScPP....7...24C,PhysRevX.9.031009, PhysRevB.98.205136, PhysRevB.99.224307, PhysRevB.100.134306, choi2019quantum, PhysRevB.100.064204, gullans2019dynamical, PhysRevResearch.2.013022,2019arXiv190804305B,jian2019measurementinduced,2019arXiv191000020G, 2019arXiv191100008Z,zhang2020nonuniversal}. As a function of the rate of measurements, the entanglement entropy of individual quantum trajectories goes from volume-law to area-law entanglement scaling: enough local measurements can collapse the many-body wavefunction into an area-law entangled state. Numerical studies in 1+1d indicate that this transition is continuous with emergent conformal invariance at the critical point~\cite{PhysRevB.100.134306}. A closely-related transition was proposed earlier by tuning the bond dimension of a state obtained at the boundary of a two-dimensional random tensor network~\cite{Hayden2016, PhysRevB.100.134203}. In all such instances the entanglement entropy in the scaling limit takes the universal form $S-S_c=F((g-g_c)L^{1/\nu})$, with $g$ the parameter driving the transition -- either the bond dimension $D$ in the case of random tensor networks, or the measurement rate $p$ for random quantum circuits -- and $S_c$ the entanglement entropy at criticality.

Theories of such entanglement transitions have been proposed using a replica approach for both random tensor networks~\cite{PhysRevB.100.134203}, and (Haar) random quantum circuits~\cite{PhysRevX.7.031016, PhysRevX.8.021014, PhysRevX.8.021013, PhysRevX.8.041019,PhysRevLett.121.060601,PhysRevX.8.031058, PhysRevB.99.174205, PhysRevX.8.031057, PhysRevLett.123.210603} combined with generalized measurements~\cite{2019arXiv190804305B,jian2019measurementinduced}. The calculation of the entanglement entropies in such circuits/networks can then be mapped onto a two-dimensional statistical mechanics model: the area- to volume-law entanglement transition corresponds to an ordering transition in the statistical mechanics model~\cite{PhysRevB.100.134203,2019arXiv190804305B,jian2019measurementinduced}. While this approach explains the general scaling properties of entanglement transitions, the resulting statistical mechanics models cannot be solved in the replica limit except in some special cases. Computing the scaling properties and the critical exponents of entanglement transitions remains a formidable challenge. 

In this letter, we propose a {\em mean-field theory} of entanglement transitions by studying random tree tensor networks. (We expect that related transitions can also be realized in certain models of random unitary dynamics with projective measurements~\cite{private-comm-HusDav}.)  As a function of bond dimension $D$, we show that random tree tensor network wavefunctions go from area-law to logarithmic entanglement scaling. The calculation of the entanglement entropy maps exactly onto a replica statistical mechanics model defined on a {\em Cayley tree}, which we argue has mean-field-like behavior.  This allows us to study this phase transition in detail using the so-called cavity method~\cite{mezard1987spin}. Remarkably, the absence of loops on the Cayley tree allows us to analyze analytically the universality class of this transition in the replica limit. 
We verify our predictions numerically by working directly with the quantum states defined as random tree tensor networks.

\begin{figure}
\includegraphics[width=1\linewidth, trim = 0in 4.5in 0in 0.5in]{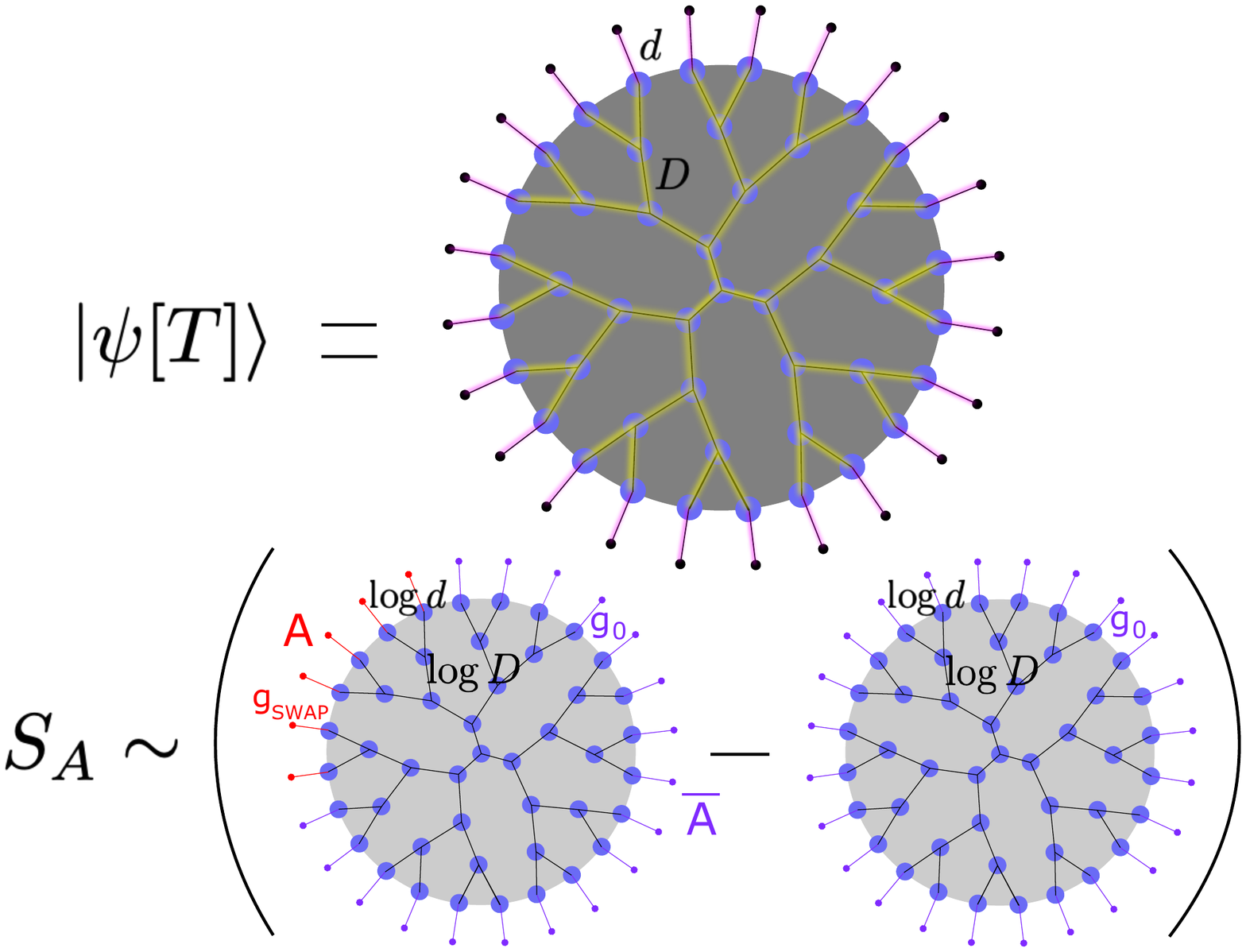}
\caption{{\bf  Random tree tensor networks.} Top: tree tensor network geometry: the physical quantum degrees of freedom live at the boundary (``leaves'') of the tree. Bottom: the entanglement entropy of a region $A$ at the boundary can be expressed as the free energy cost of a domain wall of a classical statistical mechanics model defined on the Cayley tree. }
\label{Fig: Cayley_tree}
\end{figure}

\textit{Random tree tensor networks ---} We consider one-dimensional quantum wavefunctions $\Ket{\psi}$ given by tree tensor networks (Fig.~\ref{Fig: Cayley_tree}) --- see Refs.~\cite{PhysRevA.74.022320,PhysRevB.82.205105,RevModPhys.82.277,ORUS2014117} and references within. The physical degrees of freedom are qudits of dimension $d$, which live at the boundary of the tree tensor network. Let $q$ be the coordination number of the tree, and $D$ the bond dimension of the tensor network. We choose the tensors to be random~\cite{Hayden2016}, obtained by drawing the tensor for each node of the tree independently from a featureless Gaussian distribution characterized by zero mean and unit variance. Because of the tree geometry, such wavefunctions can have logarithmic entanglement scaling, contrary to matrix-product states for example. 

Our main goal is to study the entanglement properties of wavefunctions generated from this random ensemble. This approach is inspired in spirit by random matrix theory, but it allows us to include some locality structure in the geometry of the ``bulk'' tensor network, controlling the entanglement of the boundary physical system.  
We will focus on the tensor-averaged Renyi entropies 
\begin{equation} \label{Eq: EE}
S_A^{(n)}= \frac{1}{1-n} \overline{\log \frac{{\rm tr} \rho_A^n}{({\rm tr} \rho)^n}},
\end{equation}
where $\overline{(\dots)}$ refers to averaging over random tensors, and $\rho_A$ is the reduced density matrix in some contiguous interval $A$ of size $L_A$ obtained from tracing out the complement of $A$ in $\rho = \Ket{\psi}\Bra{\psi}$ (Fig.~\ref{Fig: Cayley_tree}). 

\textit{Statistical Mechanics model ---} In order to compute these Renyi entropies, we follow Refs.~\cite{PhysRevB.99.174205,PhysRevB.100.134203,2019arXiv190804305B,jian2019measurementinduced} and use a replica trick $\log {\rm tr} \rho_A^n = \lim_{m \to 0} (({\rm tr} \rho_A^n )^m-1)/m$. This allows us to express~\eqref{Eq: EE} as 
\begin{equation} \label{Eq: EEstatmec}
S_A^{(n)}=\frac{1}{n-1}  \lim_{m\to 0}\frac{1}{m} \left( {\cal F}_A-{\cal F}_0 \right),
\end{equation}
with $\mathcal{F}_{A,0} = -\log \mathcal{Z}_{A,0}$ and $\mathcal{Z}_0 \equiv \overline{(\text{tr} \rho^n)^m}$, $\mathcal{Z}_A \equiv \overline{(\text{tr} \rho_A^n)^m}$. Using this exact identity, the calculation of the Renyi entropies reduces to computing $\mathcal{Z}_0$ and $\mathcal{Z}_A$, and to evaluate the replica limit~\eqref{Eq: EEstatmec}. When $m$ and $n$ are integers, the averages in $\mathcal{Z}_0$ and $\mathcal{Z}_A$ can be evaluated analytically using Wick's theorem.
One can then express the  partition functions $\mathcal{Z}_{A}$ and $\mathcal{Z}_{0}$ in terms of a classical statistical mechanics model, whose degrees of freedom are {\em permutations} $g_i \in S_{Q=nm}$ labelling different Wick contractions at each vertex of the tensor networks. 
Since the degrees of freedom of this statistical mechanics model live on the nodes of the tree tensor network, they form a {\em Cayley tree}, and $\mathcal{Z}_{A}$ and $\mathcal{Z}_{0}$ differ only in their boundary conditions. Using the results of Ref. \onlinecite{PhysRevB.100.134203}, we find that ${\cal Z}_0 = \sum_{ \lbrace g_i \rbrace} {\rm e}^{-{\cal H}}$, with the following nearest-neighbor Hamiltonian
\begin{equation} \label{Eq: model}
{\cal H}=-\sum_{\braket{i,j}}J_{\braket{i,j}}C(g_i^{-1}g_{j}),
\end{equation} 
where $C(g)$ counts the number of cycles in the permutation $g$, $J_{\braket{i,j}}=\log D$ with $D$ the bond dimension for links connecting bulk tensors, and  $J_{\braket{i,j}}=\log d$ (with $d$ the dimension of the Hilbert space of the boundary physical qudits) for boundary couplings involving physical degrees of freedom. This Hamiltonian is invariant under global left/right multiplication of the degrees of freedom $g_i$ by any permutation $h \in S_Q$, so it has a $S_Q \times S_Q$ symmetry. In this mapping, the trace over physical degrees of freedom in  $\mathcal{Z}_0 = \overline{(\text{tr} \rho^n)^m}$ forces the permutations on the boundary sites corresponding to the physical qudits to be fixed to the identity permutation $g_\partial= g_0= ()$ in ${\cal Z}_0$. Meanwhile, boundary permutations in ${\cal Z}_A$ are fixed to identity if they belong to $\overline{A}$ (the complement of A), whereas they are fixed to a different permutation $g_{\rm SWAP} = (12 \dots n)^{\otimes m}$ if they belong to $A$. The permutation $g_{\rm SWAP}$ arises from enforcing the partial trace in  $\mathcal{Z}_A \equiv \overline{(\text{tr} \rho_A^n)^m}$. Note that $C(g)$ is maximum for the identity permutation, so that the Hamiltonian~\eqref{Eq: model} corresponds to ferromagnetic interactions.

In the language of this statistical mechanics model, the Renyi entropies~\eqref{Eq: EEstatmec} can be computed from the free energy cost of inserting a {\em domain wall} between the boundary permutations $g_0$ and $g_{\rm SWAP}$ at the entanglement interval. This provides a very simple picture of the scaling of the entanglement entropy as a function~\footnote{While the bond dimension $D$ is in principle an integer, it is possible to construct tensor networks using projected entangled pairs that correspond to arbitrary $D$. In the following, we will assume $D >1$ to be a real number.} of bond dimension $D$. If $D$ is small (near 1), we expect the statistical mechanics model~\eqref{Eq: model} to be disordered (paramagnetic), and the free energy cost in~\eqref{Eq: EEstatmec} will not scale with $L_A$: this corresponds to an area-law phase. If on the other hand $D$ is large, the statistical mechanics model is in an ordered (ferromagnetic) phase with all bulk permutations aligned and equal to $g_0$, and the free energy cost in~\eqref{Eq: EEstatmec} will be given by the energy penalty of the bonds frustrated by the domain wall minimizing this energy (``minimal cut'' through the network). For large $L_A$ and generic intervals $A$, this minimal domain wall cuts $\sim \log L_A$ bonds of the tensor network (Cayley tree) corresponding to logarithmic entanglement scaling $S_A \sim (\log D) \log L_A$. 
This implies that the ordering transition of~\eqref{Eq: model} at a critical coupling $J_c = \log D_c$ corresponds to an area- to logarithmic scaling of the Renyi entropies of the random tree tensor networks. 

\textit{$Q=2$ replicas and cavity method ---} In order to gain some insight into the scaling of the entanglement entropy, we start by analyzing the simpler case of $Q=2$ replicas. As we will argue below, the mean-field nature of the statistical mechanics model on the Cayley tree will make critical properties mostly independent of $Q$, allowing us to generalize this insight to the replica limit $Q \to 0$. For $Q=2$, eq.~\eqref{Eq: model} is simply an Ising model. If we let $g_i=\pm 1$ be the two elements of $S_2 \cong \mathbb{Z}_2$, \eqref{Eq: model} reads ${\cal H}=-\sum_{\braket{i,j}}J_{\braket{i,j}} (3+g_i g_j)/2$, which up to an irrelevant additive constant, is an Ising model with coupling $K = J/2 = (\log D)/2$. To proceed, we use the so-called \textit{Cavity Method}~\cite{mezard1987spin, mezard2009information,RevModPhys.80.1275} which is a standard approach for solving statistical mechanics problems on tree-like graphs. We start from an Ising model with coupling $K$, and generic boundary fields $h_i$ acting on the boundary sites of the Cayley tree.
It is straightforward to show that all boundary spins can be decimated, at the price of introducing new effective fields acting on the next layer of the tree, which now forms the new boundary. This process can then be iterated, and the resulting recursion (``cavity'') equations for uniform boundary fields are then given by $\left[\sum_{\sigma_i=\pm 1} \exp(K \sigma_i \sigma_j+h^{(k+1)} \sigma_i)\right]^{q-1}=\mathcal{C}\exp(h^{(k)}\sigma_j)$, for some constant $\mathcal{C}$. Here we have assumed that we are working with ${\cal Z}_0$ for simplicity so that the boundary fields are uniform, but this approach can be readily extended to arbitrary inhomogeneous boundary fields.The critical behavior of this model is then easily deduced from solving for the cavity fields recursively~\cite{SupMat}. Approaching the transition from the paramagnetic phase, we find that the magnetization at the root of the tree decays exponentially with the number of layers $N$, $\braket{\sigma_0}\sim \exp(-N/\xi)$ with a correlation length $\xi =-1/\log((q-1)\tanh K)$ that diverges at the critical coupling $K_c=\text{arctanh}(1/(q-1))$, which is finite for coordination number $q >1$. Expanding near the critical point yields $\xi\sim |K-K_c|^{-\nu}$, with $\nu=1$. On the ferromagnetic side, we have $\braket{\sigma_0}\sim h\sim (K-K_c)^{\beta}$, with $\beta=1/2$.
(A procedure to access this exponent was proposed in Ref.~\cite{2019arXiv191000020G} in the context of random circuits.) The order parameter exponent $\beta = 1/2$ takes the mean-field value for a transition in the Ising universality class, a general feature of statistical mechanics on the Cayley tree~\cite{Katsura74}. On the other hand, the correlation length exponent $\nu=1$ is inherited from quasi-one-dimensional physics,  as has been noted previously~\cite{huExactCorrelationFunctions1998}.

\begin{figure}
\centering
\includegraphics[width=1 \linewidth]{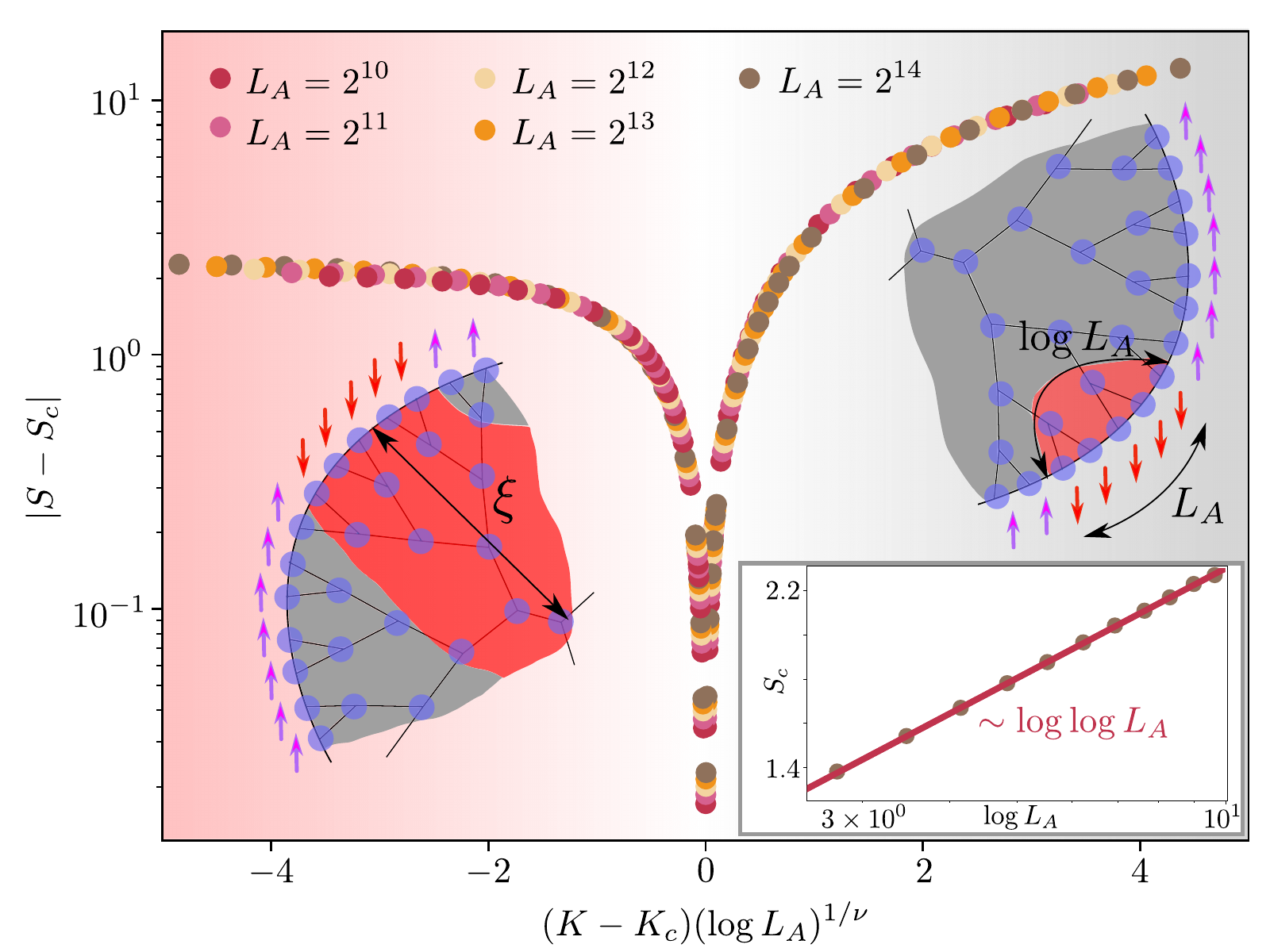}
\caption{{\bf Entanglement scaling.} Collapse of the boundary domain wall free energy cost for $Q=2$ replicas, as a proxy for the entanglement entropy in the replica limit $Q \to 0$. For $K=(\log D)/2>K_c$ the domain wall mostly follows a minimal cut through the bulk, so its energy scales logarithmically with the interval size $L_A$. For $K<K_c$, the domain wall fluctuates through the bulk over a number of layers given by the correlation length, which diverges as $\xi \sim |K-K_c|^{-\nu}$ with $\nu=1$. Inset: at criticality, the entanglement scales as $S \sim \log \log L_A$. }
\label{Fig: collapse}
\end{figure}

\textit{Entanglement Scaling ---} The cavity method above can readily be applied to arbitrary configurations of the boundary fields, and can be used to evaluate eq.~\eqref{Eq: EEstatmec} in the case of $Q=2$ replicas~\cite{SupMat}. We denote the averaged free energy cost of a domain wall $S(L_A) = \mathcal{F}_A -\mathcal{F}_0$, which is the quantity which becomes entanglement entropy in the limit $Q \to 0$ from eq.~\eqref{Eq: EEstatmec}. 
On the paramagnetic side of the transition (small $K=(\log D) /2$), the Ising order decays $\xi$ layers into the bulk, so we expect the entanglement to saturate to a constant value $S(L_A \to \infty) \propto \log \xi$, corresponding to area law scaling. This is consistent with our numerical results~\cite{SupMat}, which indicate a divergence $S(L_A \to \infty) \sim - \alpha\log (K_c - K)$ as $K\rightarrow K_c^-$. The saturation to this area law value occurs for $L_A \gg \xi_\star$ with the crossover scale $\xi_\star = {\rm e}^{\xi} = {\rm e}^{C/|K-K_c|}$. Therefore, while $\nu=1$ in the bulk, in terms of the entanglement scaling the relevant diverging length scale diverges exponentially near the transition, due to the tree geometry. On the ordered side of the transition ($K>K_c$), $S(L_A)$ is proportional to the energy cost of the domain wall which scales as the number of layers into the bulk $\sim \log L_A$. As expected from general scaling arguments, the prefactor is set by $\xi$, and we find $S(L_A) \sim  \frac{\log L_A}{\xi}$. Finally, scaling at the critical point is required to be $ S(L_A) \sim \alpha \log \log L_A$ by general scaling considerations from the behavior in the phases, in good agreement with our numerical solution to the cavity equations for the system sizes we can access (Fig.~\ref{Fig: collapse}).
 In summary, we have
\begin{align}\label{Eq: Entropy}
S \sim
\begin{cases}
\frac{\log L_A}{\xi}+ \alpha \log\log L_A, & K\rightarrow K_c^+,  \\
\alpha\log \log L_A, & K=K_c,\\
\alpha \log \xi, & K\rightarrow K_c^-.
\end{cases}
\end{align}
%

We find that our results are consistent with the entanglement scaling at entanglement transitions in quantum chaotic systems subject to projective measurements or in wavefunctions given by square random tensor networks upon replacing $\log L_A \rightarrow L_A$~\cite{PhysRevX.9.031009, PhysRevB.100.134203, PhysRevB.100.134306}. This is because geodesics (minimal cut minimizing the domain wall energy at large bond dimension) in flat 2D Euclidean space are given by straight lines,  whereas they scale with the logarithm of the size of region $A$ on the Cayley tree. These different regimes can be summarized by the universal scaling form $S-S_c=F((K-K_c)(\log L_A)^{1/\nu})$ with $\nu=1$ shown in Fig.~\ref{Fig: collapse}.

\textit{Replica limit ---} So far our results for the bulk critical exponent and for the entanglement scaling~\eqref{Eq: Entropy} were inferred from the case of $Q=2$ replicas for simplicity.  We now discuss how one can obtain the critical properties in the replica limit $Q \to 0$ of eq.~\eqref{Eq: EEstatmec}. It is possible~\cite{SupMat} to apply the cavity method to the model~\eqref{Eq: model}, but the number of cavity fields is then given by the number of irreducible representations of $S_Q$. As a result, the replica limit $Q \to 0$ is still out of reach on the Cayley tree. 
To proceed, we use the following trick based on universality: we modify the Boltzmann weights of the model eq.~\eqref{Eq: model} while preserving the $S_Q \times S_Q$ symmetry of the Hamiltonian~\eqref{Eq: model}. 
Therefore, we introduce a different statistical mechanics model
\begin{equation} \label{Eq:ModifiedModel}
{\cal H}_{\rm modified} = - \sum_{\langle i,j \rangle} \log \left( 1 + K {\bar \chi}(g_i^{-1} g_j)\right),
\end{equation} 
where ${\bar \chi}(g)=\frac{Q-1}{Q!} \chi(g)$ with $\chi$  the character of the {\em standard} representation of the symmetric group $S_Q$. This model is still invariant under left/right multiplication by elements of $S_Q$, and since the standard representation is faithful and well-defined for any $Q$, we do not expect this modified model to have an enlarged symmetry. (This is inspired by the $O(N)$ model, whose critical behavior in 2D was understood by Nienhuis~\cite{PhysRevLett.49.1062} by introducing a different model with the same symmetry group.)

\begin{figure}
\centering
\includegraphics[width=1 \linewidth]{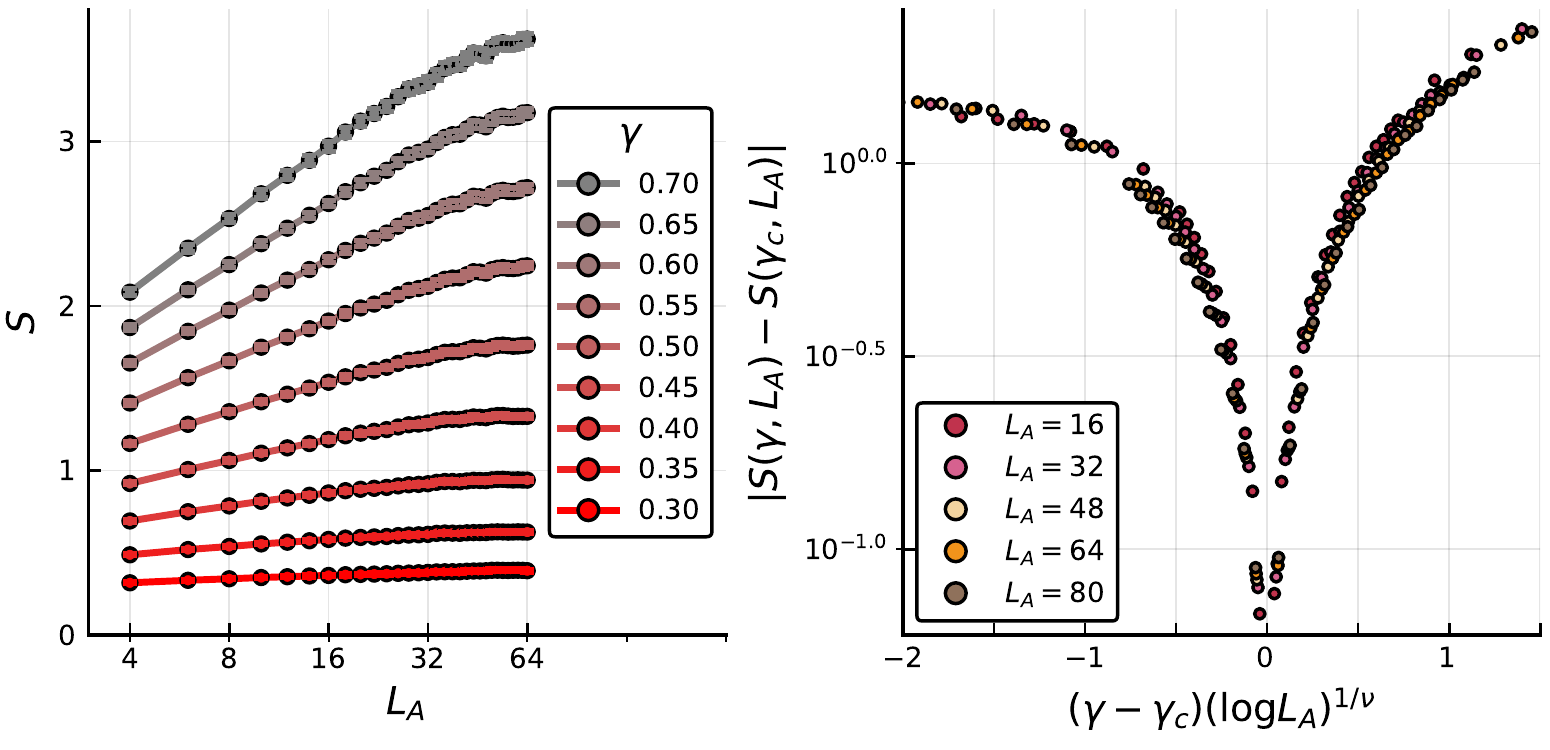}
\caption{{\bf Numerical results.} Left panel: averaged von Neumann entropy for random tree tensor network states of size $L=256$ as a function of the subsystem size $L_A$ for various values of $\gamma$, where $\gamma \in [0,1]$ is a  parameter tuning continuously the bond dimension between $D=1$ and $D=3$ (see text). Right panel: collapse of the data with $\gamma_c = 0.47$ and $\nu = 1$.}
\label{Fig: numericsmain}
\end{figure}

Remarkably, for uniform boundary conditions $g_\partial = g_0 = ()$ (corresponding to ${\cal Z}_0$),  the modified model~\eqref{Eq:ModifiedModel} is still solvable on the Cayley tree with coordination number $q=3$ using a single cavity equation for any $Q$. The cavity equation reads $\sum_{g_i} (1+ h^{(k)} {\bar \chi}(g_i))^2 (1+K {\bar \chi}(g_i^{-1} g_j)) = C (1+h^{(k-1)} {\bar \chi}(g_j))$. 
Using standard representation theory results, 
we find the following recursion relation for the boundary cavity fields
\begin{equation} \label{Eq:ModifiedModelCavity}
h^{(k-1)} = \frac{K}{Q!} \frac{2h^{(k)} +(h^{(k)})^2 \frac{Q-1}{Q!}}{1+ (h^{(k)})^2 \left(\frac{Q-1}{Q!} \right)^2}.
\end{equation} 
We can now analytically continue $Q$ in this equation, and study the critical behavior as a function of $Q$. We analyzed the fixed points of this recursion relation and their stability as a function of $Q$. For $Q>1$, we find first order transitions (with $Q=2$ being special), while for $Q<1$ there is a second order transition for $K_c = Q!/2$. For $K<K_c$, the correlation length reads $\xi^{-1}= \log \frac{K_c}{K}$ so we find $\nu=1$ as in the Ising ($Q=2$) case. For $K>K_c$, the cavity fields flow to a non-zero value which scales as $\sim (K-K_c )$, so we find $\beta=1$ which is the mean-field magnetization exponent of the $n$-state Potts model with $n < 2$. In the replica limit, we thus find $\nu=\beta=1$, which coincide with the critical exponents of the $n$-state Potts model on the Cayley tree (for $n<2$). Those exponents {\em do not} depend on the replica number $Q$, as expected from mean-field critical behavior in general --- the only exception is the exponent $\beta$ which happens to be different for $Q=2$ for symmetry reasons. We expect these exponents to control the critical behavior of our model~\eqref{Eq: model} in the replica limit $Q \to 0$, and while we unfortunately cannot solve the modified model~\eqref{Eq:ModifiedModelCavity} with inhomogeneous boundary conditions, we also expect the general scaling~\eqref{Eq: Entropy} with $\nu=1$ to hold for $Q \to 0$. 

{\textit{Numerical results --- } We verify our predictions by generating tree tensor network states and computing their entanglement properties numerically. Each state consists of random, gaussian-distributed tensors of dimension $D$ on each node of the Cayley tree. By tuning the bond dimension we find a phase transition from area-law to logarithmic scaling of the entanglement entropy, with $D=1$ (trivially) showing area-law scaling and $D=3$ showing clear logarithmic scaling.
As the dimension of tensors must be integer, we augment these states with additional tensors on each bond of the tree to interpolate between integer bond dimensions $D=1$ and $D=3$. With the size of the tensors on the nodes fixed at $D=3$, we insert on each bond diagonal tensors with elements $\left(1, \gamma, \gamma^2 \right)$, with the parameter $\gamma$ tuned continuously from $\gamma=0$, corresponding to $D=1$, to $\gamma=1$, corresponding to $D=3$ (see~\cite{SupMat}). Upon tuning $\gamma$, we find a phase transition from area-law to logarithmic scaling of the entanglement entropy (Fig.~\ref{Fig: numericsmain}), consistent with the mean-field theory results detailed above. We estimate the location of the critical point $\gamma_c$ to be in the interval $[0.4, 0.6]$ and the critical exponent $\nu$ to take a value in $[1, 1.5]$. The precision is limited due to the rather small depth of the Cayley tree that is accessible numerically; however, we find that the quality of the collapse improves with system size and is comparable to our results for the Ising model on equally small Cayley trees\cite{SupMat}.

\textit{Discussion --- } We have studied a new entanglement transition from area-law to logarithmic scaling of entanglement in random tree tensor networks. This transition can be analyzed using a mapping onto a replica statistical mechanics model on the Cayley tree which exhibits mean-field critical behavior. We computed exactly the critical exponents $\nu=\beta=1$ relevant to this entanglement transition, and inferred the general scaling properties of the entanglement near criticality. We checked our predictions numerically by computing directly the entanglement of random tree tensor network states. It would be interesting to find other mean-field examples of entanglement transitions, especially in the context of measurement-induced transitions in random quantum circuits. 

\begin{acknowledgments}

\emph{Acknowledgments}.---We thank Sarang Gopalakrishnan, Michael Gullans, David Huse, Sid Parameswaran and Jed Pixley  for useful discussions. R.V. also thanks Chao-Ming Jian, Andreas Ludwig, Andrew Potter and Yi-Zhuang You for collaborations on related matters.  This work was supported by the US Department of Energy, Office of Science, Basic Energy Sciences, under Early Career Award No. DE-SC0019168.  R.V. is supported by the Alfred P. Sloan Foundation through a Sloan Research Fellowship. 

\end{acknowledgments}

\bibliography{References}

%


\bigskip

\onecolumngrid
\newpage



\section*{Supplementary material}

# Supplemental Material for "Mean-field theory of entanglement transitions from random tree tensor networks"

Javier Lopez-Piqueres, Brayden Ware and Romain Vasseur[1]

[1]*Department of Physics, University of Massachusetts, Amherst, Massachusetts 01003, USA*

## SCALING OF THE ENTANGLEMENT ENTROPY AT LARGE BOND DIMENSION

In this appendix, our goal is to show that at large bond dimension, the entanglement entropy averaged over all possible contiguous entanglement regions scales as $\log L_A$, with $L_A$ the size of region $A$. (Note that there are special choices of the position of the entanglement interval $A$ for which this minimal cut could cost only $\mathcal{O}(1)$ energy.) In this regime, the underlying statistical mechanics model is effectively at zero temperature and is in a ferromagnetic phase. The free energy cost of the boundary domain wall (DW) created at the boundaries of $A$ is then purely energetic, and the problem reduces to minimizing the energy of the domain wall by minimizing the number of frustrated bonds through the bulk of the tree connecting both ends of region $A$ (see Fig. 1). This corresponds to finding the "minimal cut" through the tree tensor network.

For simplicity, we focus on the rooted Cayley tree of coordination number $q = 3$. We will also take entanglement regions $A$ of size $L_A = 2^m$ with $m$ some positive integer (as in the main text). We can recast the shortest path problem between the two ends of a boundary entanglement domain of size $L_A = 2^m$ to that of finding the shortest path from one end to the top of a tree of size $L_A = 2^m$ (see Fig. 1). The resulting path's length will be asymptotically half of the total length of the minimal cut. Consider a tree of $m$ shells/layers (excluding the root), with the leaves a the $m^{\text{th}}$ (boundary) shell:

- For every vertex at shell $n$ allocate a new vertex below it, at shell $n + 1$. Add two further vertices at each side of the root of the tree (see Fig. 1 a)). Note: the new set of vertices provide us with the location of the ends of the possible entanglement cuts. To avoid confusion we name these new vertices as *stars*.

- Starting from any star create a path to the top of the tree with the only constraint that consecutive shells are not connected. Note that this path is unique.

This procedure guarantees that the number of bonds broken from any star to either of the top stars is minimized. We note that there might be more than one possible shortest path for every star at the boundary, which translates into a degenerate ground state in the statistical mechanics problem (whose degeneracy grows with $L_A$). However, this algorithm only chooses one of the possible paths. Now we shall show that indeed the average distance, averaged over all possible entanglement cuts (or equivalently, over all paths to the top of the tree starting from the stars below) scales as $\log L_A$. Denote by $d^{(m)}$ the average distance to the top, taken over all stars from shells $\leq m$. First note that there are half as many stars at shell $m$ as there are vertices, or equivalently, half as many spots where we can locate an end of the entanglement cut. These, by construction, are connected to all shells $n \leq m - 2$, with two stars at shell $m$ connected to each star at shell $n \leq m - 2$ (see Fig. 1). Moreover, the other half of the locations where an end of the entanglement cut can be placed lies at the remaining shells $n \leq m - 1$. Thus, the sought average distance can be expressed in terms of the simple recursive relation

$$d^{(m)} = \frac{1}{2}(d^{(m-2)} + 1) + \frac{1}{2}d^{(m-1)}, \tag{1}$$

where the 1 inside the parenthesis stems from the broken bond to connect shell $m$ with any shell $n \leq m - 2$. The solution for arbitrary $m$ is

$$d^{(m)} = a\left(-\frac{1}{2}\right)^m + \frac{m}{3} + \mathcal{O}(1). \tag{2}$$

for some constant $a$. The first term dies off for large $m$ and the third term is an irrelevant constant. The entanglement entropy as $K \to \infty$ for an entanglement cut of size $L_A$ thus behaves as

$$\left.\frac{1}{K}S(L_A)\right|_{K,L_A\to\infty} \sim \frac{4}{3\log 2}\log L_A, \tag{3}$$

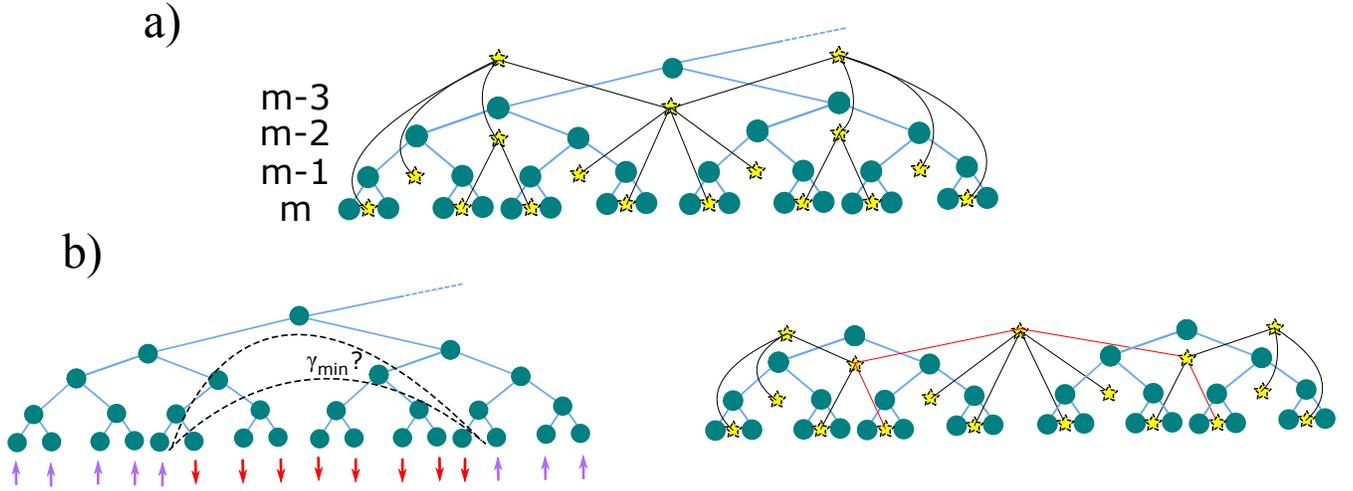

FIG. 1: a) Algorithm that finds the shortest path to the top of a tree of size $L_A = 2^4$ from any star. b) The solution to the shortest path connecting two ends of an entanglement region of size $L_A = 2^m$ can be approximated by twice the shortest path to the top of a tree of size $L = 2^m$ (right figure, shortest path in red).

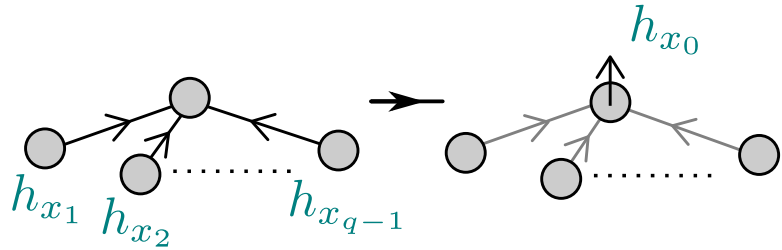

FIG. 2: Given some local fields around a root $x_0$, one can *decimate* these to obtain a new, *cavity* field at $x_0$.

where a factor of 2 comes from approximating the shortest path connecting the two ends of the entanglement cut by doubling the distance to the top of the tree, and another factor of 2 simply corresponds to the energy cost per broken bond. This coefficient is consistent with our numerical results, which show a prefactor for the logarithm of $\sim 1.9 - 2.1$ for the accessible values of $K$ and $L_A$.

### CAVITY METHOD/ BELIEF-PROPAGATION ALGORITHM

For concreteness, let us consider the Ising model on a Cayley tree $G_N$ with $N$ shells and coordination number $q$ with some arbitrary pinning fields at the leaves (boundary). One can integrate out the spins from the leaves and obtain an effective partition function on a Cayley tree with $N-1$ shells, $G_{N-1}$, with new effective fields at the leaves

$$\sum_{\{\sigma_{x_i}\}} \exp\left(K \sum_{i=1}^{q-1} \sigma_{x_i}\sigma_{x_0} + \sum_{i=1}^{q-1} h_{x_i}\sigma_{x_i}\right) = \mathcal{C} \exp(h_{x_0}\sigma_{x_0}) \quad (4)$$

These new fields are the *cavity* fields, see Fig. 2. Using the relation $\text{arctanh}(x) = \frac{1}{2}\log\left(\frac{1+x}{1-x}\right)$, we obtain the cavity fields at shell $N-1$: $h_{x_0} = \sum_{i=1}^{q-1} \text{arctanh}(\tanh K \tanh h_{x_i})$, which depend on those at shell $N$. The constant $\mathcal{C}$ is given by $\mathcal{C} = \exp(-h_{x_0}) \prod_{i=1}^{q-1} 2\cosh(K + h_{x_i})$. Iterating this procedure, one can obtain the full partition function which depends only on the interaction constant between nearest-neighbor spins, $K$, and the cavity fields from all vertices. This simple *decimation-like* picture carries on to any tree-like graph, with any number of variables at each vertex. To illustrate how critical behavior on such graphs emerges from this method, let us consider the $n$-Potts model on

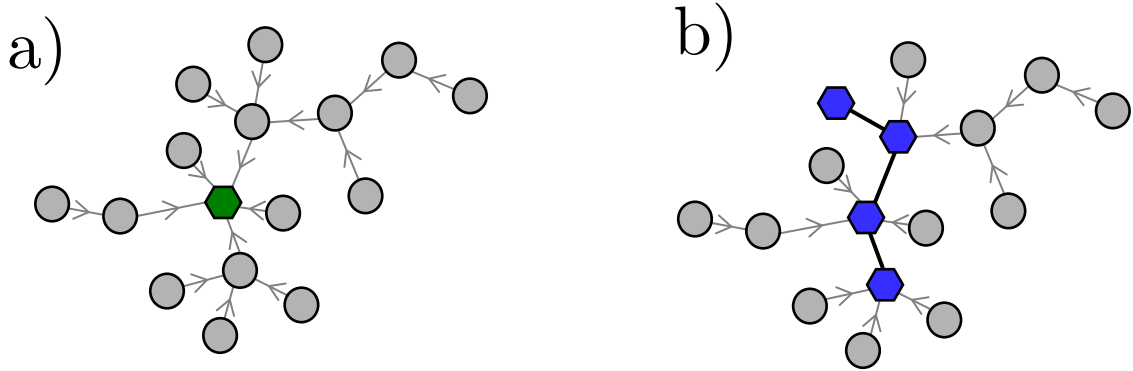

FIG. 3: a) The BP algorithm allows us to compute the local magnetization at a vertex (in green) by repeating recursively the decimating step in Fig. 2. b) One can also compute the two-point function of two spins which will depend only on the cavity fields pointing towards these.

a tree-like graph $G$ with arbitrary fields on a subset (which for simplicity we take them to be located at the set of external vertices $\partial G$). The partition function is $\mathcal{Z} = \sum_{\{s\}} \exp(-\mathcal{H}(\{s\}))$, with

$$\mathcal{H}(\{s\}) = - \sum_{\langle v,v'\rangle \subset G} K(2\delta_{s_v,s_{v'}} - 1) - \sum_{v \in \partial G} h_v(2\delta_{s_v,1} - 1) \tag{5}$$

and $s_v = 1, 2, ..., n$. Computing the partition function naively is obviously exponentially hard, with $\mathcal{O}(n^{|G|})$ steps. The Belief-Propagation (BP) algorithm performs such a sum in a smart fashion by introducing a set of *beliefs* $\nu_v(s_v)$ at each vertex $v \in G$. These are marginal distributions obtained after *decimating* a subset of spins connected to $s_v$, called *children* or *nearest descendants* of vertex $v$, hereafter denoted by $N(v)$. Because of the local nature of these distributions, we may label them by local fields: the cavity fields mentioned above. The end result is a means to carry out that sum in $\mathcal{O}(|G|)$ steps. The BP algorithm boils down to the simple update-rules

$$\nu_v(s_v) = \sum_{\{s_{v'}\}} \prod_{v' \in N(v)} \mu(s_v, s_{v'}) \cdot \nu_{v'}(s_{v'}), \tag{6}$$

where $\mu(s_v, s_{v'})$ is here playing the role of the Boltzmann weight. Note that these update-rules were used already in Eq. (4) for Ising variables, but the approach is very general. With this scheme one can evaluate very efficiently local quantities as well as correlators, see Fig. 3. As a byproduct of the BP algorithm, the critical properties of the system can be readily read off from the cavity fields.

Let us illustrate this approach in our example of the Potts model. (We will extend it to our model of interest below.) First, we associate a direction on the graph $G$, with the only constraint that the arrows point away from the external vertices and that they converge to an arbitrary vertex $v_0$, which will be the root (see Fig. 3 for an example). The *bulk* Boltzmann weights are given by $\mu(s_v, s_{v'}) \propto \exp(K(2\delta_{s_v,s_{v'}} - 1))$ and the beliefs by $\nu_v(s_v) \propto \exp(h_v(2\delta_{s_v,1} - 1))$. Given a set of boundary fields on $\partial G$, we apply Eq. (6) recursively, or explicitly

$$\mathcal{C} \exp(h_v(2\delta_{s_v,1} - 1)) = \prod_{v' \in N(v)} \sum_{t=1,...,n} \exp(K(2\delta_{s_v,t} - 1)) \exp(h_{v'}(2\delta_{t,1} - 1)), \tag{7}$$

where $N(v)$ denotes the nearest neighbor descendants of vertex $v$, that is, the set of nearest neighbors pointing towards $v$. The recursive relations yield the following equations for the cavity fields

$$h_v = \frac{1}{2} \sum_{v' \in N(v)} \log\left(\frac{2\sinh(K + h_{v'}) + ne^{-K-h_{v'}}}{2\cosh(K - h_{v'}) + (n-2)e^{-K-h_{v'}}}\right). \tag{8}$$

Setting $n = 2$ we recover the cavity field equations used in the main text for the Ising model $h_v = \sum_{v' \in N(v)} \operatorname{arctanh}(\tanh K \tanh h_{v'})$. The partition function can be obtained from $\mathcal{C}$, when scaling up the BP equations

to the whole graph

$$\mathcal{Z} = \frac{2\sinh(h_{v_0}) + ne^{-h_{v_0}}}{2\sinh(K + h_{v_0}) + ne^{-K-h_{v_0}}} \exp\left\{\sum_{v\in G} -h_v + \log\left(2\sinh(K+h_v) + ne^{-K-h_v}\right) + \sum_{v\in\partial G} h_v\right\}. \quad (9)$$

Let $v_0$ be the root of the tree-graph $G$ which is $N$ steps away from all leaves, which are pinned to a given spin/color. In this setup one expects a disordering/ordering phase transition at a temperature $T_c = K_c^{-1}$. Indeed, we expect that at high-enough temperatures colors are uncorrelated, while at low-enough temperatures the bulk of the tree takes the same colors as the boundary. In the limit $T \to T_c^+$, we expect exponentially decaying correlations characterized by the critical exponent $\nu$, whose value can be straightforwardly extracted from the cavity equations. Indeed, expanding Eq. (8) around $h_{v'} = 0$, $\forall v' \in G \setminus \partial G$ we have

$$h_{v_0} \approx \left(\frac{2(-1+e^{2K})}{-1+e^{2K}+n}\right)\left(h_{v_1} + h_{v_2} + ... + h_{v_{|N(v)|}}\right), \quad (10)$$

where we take $v_i \in N(v_0)$. We can iterate this procedure to the whole tree, descending from the root of the tree to the leaves, to arrive at

$$h_{v_0} \approx \left(\frac{2(-1+e^{2K})}{-1+e^{2K}+n}\right)^N |\partial G| \times h_{\partial G}. \quad (11)$$

We can further approximate this by taking the number of leaves of the tree to be $|\partial G| \sim \bar{\lambda}^N$, with $\bar{\lambda}$ the average degree *from the edge-perspective*[1]. Thus, in the limit $K \to K_c^-$ we have $\langle \delta_{s_{v_0,1}} - \frac{1}{n}\rangle \sim h_{v_0} \sim \exp(-N/\xi)$, with the correlation length

$$\xi = -\frac{1}{\log\left(\bar{\lambda}\frac{-1+e^{2K}}{-1+e^{2K}+n}\right)}, \quad (12)$$

which diverges at the critical value

$$K_c = \frac{1}{2}\log\left(\frac{-1+\bar{\lambda}+n}{-1+\bar{\lambda}}\right). \quad (13)$$

Expanding the correlation length around the critical point yields

$$\xi \sim |K - K_c|^{-\nu}, \quad (14)$$

with $\nu = 1$, as expected (see main text). Solving instead the fixed-point equation for the cavity field, we can also obtain the magnetization exponent $\beta$. The order parameter, which scales as $m \sim h$ with $h$ the fixed-point solution, is discontinuous at the critical point for $n > 2$ with a jump of order $\sim \log(n-1)$, corresponding to a first order transition. Right at $n = 2$ (Ising model), $m \sim \sqrt{K - K_c}$ for $K > K_c$, so that $\beta = 1/2$ while for $n < 2$ we obtain $\beta = 1$ since $m \sim (K - K_c)$. This analysis naturally applies to the Cayley tree, where the results become *exact*, where now $N(v) = q - 1$, $\forall v \in G \setminus \partial G$. We anticipate already that based on the analysis for the $n$-state Potts model, we expect the following two general features to carry over to any statistical mechanics model on tree graphs. First, as can be seen through the lens of the cavity fields, correlations will display a one dimensional character, thus we expect $\nu = 1$ for most models. More generally *small-world* graphs (such as the Cayley tree) have a relevant length-scale that scales with the log of the number of points in the graph, which justifies the expectation for the replacement of $L_A \to \log L_A$ in the expressions for the EE. (In fact, the broader class of graphs that are small-world would allow for similar scaling properties of the EE found in this work, albeit possibly with a different critical exponent $\nu = 1/2$ [2]).

## ENTANGLEMENT SCALING FROM THE $Q = 2$ REPLICAS MODEL

We can directly compute a proxy for the entanglement entropy from the Ising model ($Q = 2$ replicas) as the free energy cost of a boundary domain wall $S = \mathcal{F}_A - \mathcal{F}_0$, with $\mathcal{F} = -\log \mathcal{Z}$. This can be evaluated from the expression (9) with $n = 2$ where $h_v = +1$ if $v \in \overline{A}$ and $h_v = -1$ if $v \in A$, such that $A \cup \overline{A} = \partial G_N$. As in the main text, we focus on a coordination number $q = 3$. Recall that as a result of the irregularity at the leaves of the Cayley tree, we must average

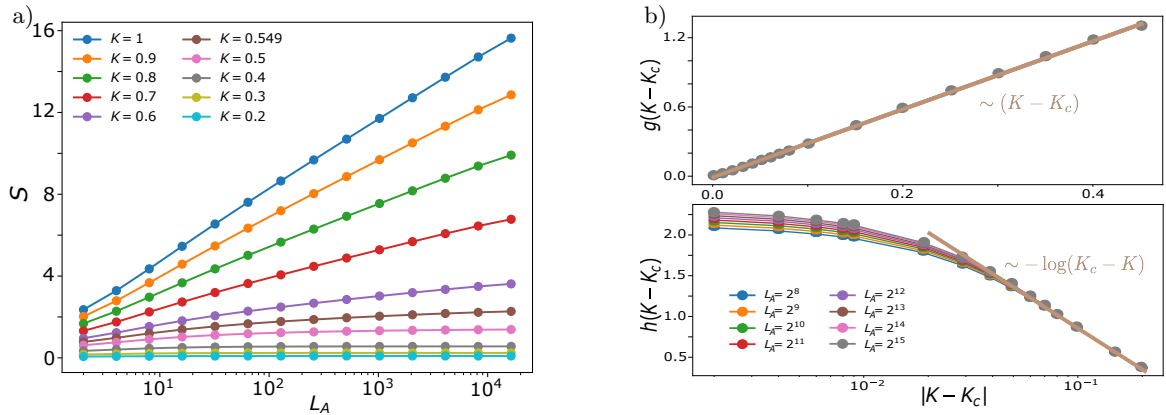

FIG. 4: a) Entanglement entropy scaling from the free energy cost of a boundary DW in the Ising model, as a function of the size $L_A$ of the boundary region $A$, for various couplings strengths $K = (\log D)/2$. b) Top: fit of the coefficient in front of the logarithm in $S - S_c = g(K - K_c) \log L_A$ in the ferromagnetic phase. Bottom: Fit of the coefficient in the paramagnetic ("area-law") phase $S = h(K - K_c)$.

the "entanglement entropy" (domain wall free energy cost) $S$ over all non-equivalent connected boundary regions $A$. The results are shown in Fig. 4. In the ferromagnetic phase, $K > K_c$, with $K_c = \operatorname{arctanh}(1/2)$, the entanglement entropy scales as $S - S_c = g(K - K_c) \log L_A$, with $g(K - K_c) \sim \xi^{-1}$, with the correlation length $\xi \sim |K - K_c|^{-1}$. In the disordered phase, $K < K_c$, the entanglement entropy scales instead as $S = h(K - K_c)$ with $h(K - K_c) \sim \log \xi$, corresponding to area-law scaling. At criticality, the entanglement entropy scales with the double logarithm of the size of region $A$, $S_c \sim \log \log L_A$. This is the only possibility at the critical point given the scaling of the entanglement entropy at both sides of the phase transition. In particular, we expect that in the area-law phase, the system is ordered only $\xi$ layers into the bulk. Due to the tree structure, this must coincide with the only relevant length-scale available, $\log L_A$. This is just a restatement of the scaling hypothesis on the Cayley tree. We confirm this expectation numerically in Fig. 5 where we show that the best fit at criticality displays a scaling of the form $S_c = C + \alpha \log \log L_A$, for some non-universal constant $C$ (for comparison we also show other fits of the form $S_c = a + b(\log L_A)^c$, showing that the best fit occurs as we take $c \to 0^+$). These different behaviors can be captured by a single universal scaling function $S - S_c = F((K - K_c)(\log L_A)^{1/\nu})$ with:

$$F(x) \sim \begin{cases} x^\nu, & x > 0, \\ 0, & x = 0, \\ -\log|x|^{-\nu}, & x < 0, \end{cases} \qquad (15)$$

with $\nu = 1$, as shown in the main text. From the random tensor network point of view, one has a bulk bond dimension corresponding to $D = \exp(2K)$, as well as a *boundary* bond dimension $d$ (see main text). This scaling is thus only valid at large enough $L_A$, as for small $L_A$ it is energetically favorable for the domain wall to cut through the $\sim L_A$ boundary links with cost $S(L_A) \sim (\log d) L_A$. We thus expect volume-law scaling for small $L_A$ such that $(\log d) L_A \ll (\log D) \frac{\log L_A}{\xi}$, crossing over to the above scaling for large intervals.

In Fig. 6 we show a comparison of the collapsed data from solving the case of $Q = 2$ replicas with the tree tensor network results, when using the same system sizes. We see that the best collapse for system sizes $L \sim 256$ actually occurs when setting $\nu \sim 1.5$ on both the classical model and on tree tensor networks. However, such a discrepancy is consistent given that the relevant length scale on trees is $\log L$. We also perform a collapse from the classical model using $\nu = 1$ which agrees with the results from Fig. 3 of the main text.

## CAVITY EQUATIONS FOR THE STATISTICAL MECHANICS MODEL DESCRIBING RANDOM TENSOR NETWORKS

Let us now turn our attention to the statistical mechanics model underlying the calculation of the entanglement entropy of random tensor networks, as described in the main text. In its most general version, the Boltzmann weights

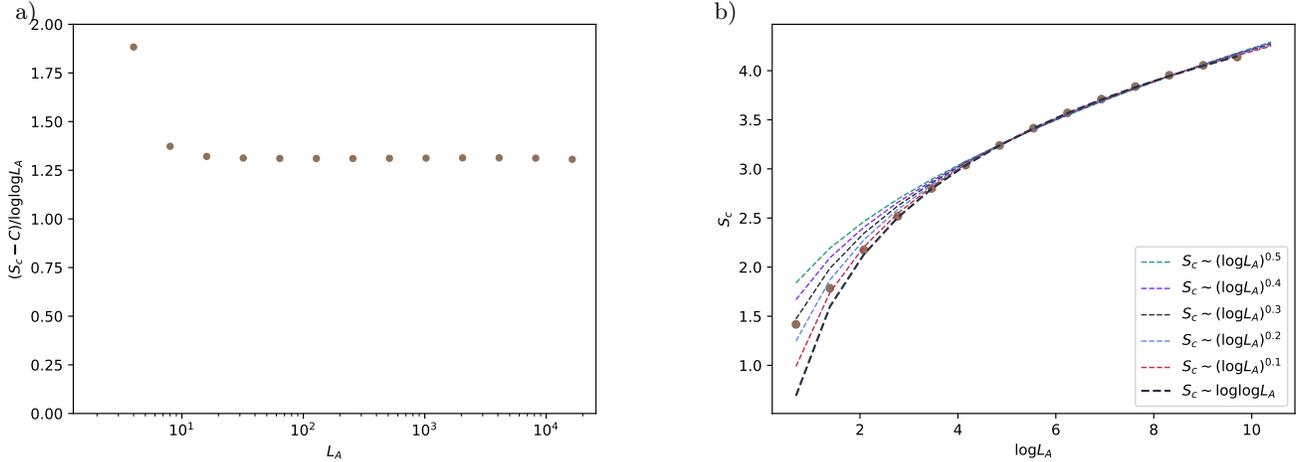

FIG. 5: Entanglement entropy scaling at $K_c$ from the free energy cost of a boundary DW in the Ising model. a) Behavior of $S_c/\log\log L_A$ follows a constant as $L_A \to \infty$. b) Comparison with other scalings of the form $S_c = a + b(\log L_A)^c$.

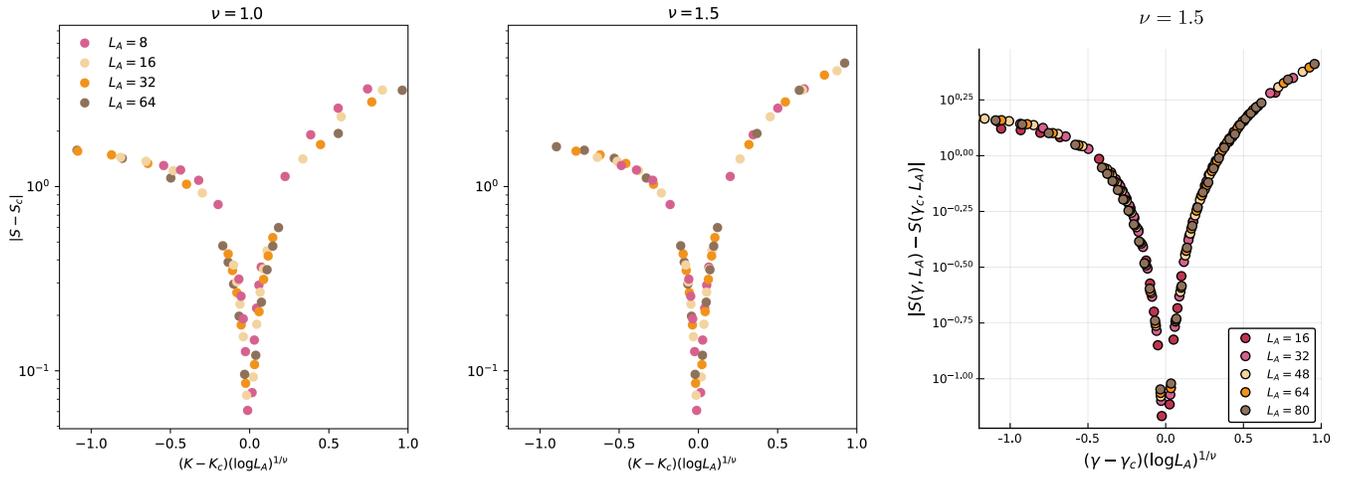

FIG. 6: Collapse for an artificially small system size $L = 256$ (matching the tensor network calculations) solving the classical statistical model with $Q = 2$ replicas (using $\nu = 1$ and $\nu = 1.5$) and the results from tree tensor networks (using $\nu = 1.5$).

can be written as

$$\mu(g_v, g_{v'}) \propto 1 + \sum_{\xi \neq 1} K_\xi \bar{\chi}_\xi(g_v^{-1} g_{v'}), \tag{16}$$

where $\xi \neq 1$ labels irreducible representations (irrep) $V_\xi$ of $S_Q$ different from the trivial one, the coupling constants $K_\xi$ depend on the irreducible representation $\xi$, $\bar{\chi}_\xi(g) = \frac{d_\xi}{Q!}\chi_\xi(g)$ with $\chi_\xi$ the character of the irrep $V_\xi$ and $d_\xi$ its dimension. Considering homogeneous pinning fields at the leaves (boundary) of the Cayley tree, the "beliefs" now read $\nu(g_v) \propto [1 + \sum_{\xi \neq 1} h_\xi^{(k)} \bar{\chi}_\xi(g_v)]^{q-1}$. Recall that, using standard representation theory results, we have $\sum_g \chi_\xi(g) = Q!\delta_{\xi,1}$, $\sum_g \chi_\xi(g)\chi_{\xi'}(g^{-1}h) = \delta_{\xi,\xi'}\frac{Q!}{d_\xi}\chi_\xi(h)$ and $\chi_{\xi \otimes \xi'}(g) = \chi_\xi(g)\chi_{\xi'}(g) = \sum_{\xi''} g_{\xi,\xi'}^{\xi''}\chi_{\xi''}(g)$, with $g_{\xi,\xi'}^{\xi''}$ the Kronecker











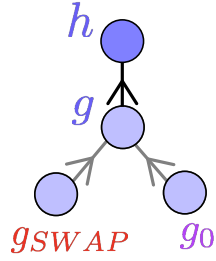

FIG. 7: Decimation of a "spin" (permutation) $g$ when introducing a DW at the boundary separating $g_{SWAP}$ and $g_0$.

coefficient (such that $V_\xi \otimes V_{\xi'} = \oplus_{\xi''} g_{\xi,\xi'}^{\xi''} V_{\xi''}$). Focusing on $q = 3$, we get the recursion relations for the cavity fields

$$h_\xi^{(k)} = \frac{K_\xi}{Q!} \frac{2h_\xi^{(k+1)} + \sum_{\xi',\xi'' \neq 1} h_{\xi'}^{(k+1)} \overline{g}_{\xi',\xi''}^{\xi} h_{\xi''}^{(k+1)}}{1 + \sum_{\xi' \neq 1} \left(h_{\xi'}^{(k+1)}\right)^2 \left(\frac{d_{\xi'}}{Q!}\right)^2}, \tag{17}$$

where the rank-3 tensor $\overline{g}_{\xi',\xi''}^{\xi} = \frac{d_\xi d_{\xi'}}{d_{\xi''}} \frac{1}{Q!} g_{\xi',\xi''}^{\xi}$. From linearizing these equations near small fields, it is clear that the correlation length critical exponent is fixed to $\nu = 1$ as in the Potts/Ising models above. Unfortunately the above equations are not easy to study for general $Q$ and to analytically continue to $Q \to 0$ since the number of irreps depends on $Q$. However, as explained in the main text, a dramatic simplification occurs if we keep only the standard irrep, that we label by $\xi = 3$: $K_\xi = K_3 \delta_{\xi,3}$. Starting from boundary fields that are also along the standard irrep, $h_\xi = h_3 \delta_{\xi,3}$, the recursion relations (17) can be reduced to a single equation for the fields along the standard irrep

$$h_3^{(k)} = \frac{K_3}{Q!} \frac{2h_3^{(k+1)} + \left(h_3^{(k+1)}\right)^2 \left(\frac{Q-1}{Q!}\right)}{1 + \left(h_3^{(k+1)}\right)^2 \left(\frac{Q-1}{Q!}\right)^2}, \tag{18}$$

where we have used that $d_3 = Q - 1$ for the standard irrep and that $g_{3,3}^3 = 1 \ \forall Q$. This equation can be analytically continued to $Q \to 0$. Solving the fixed point equation, we find the critical exponents $\nu = \beta = 1$ for all $Q < 1$. For $Q > 1$, one gets a first order phase transition — similar to the case of the Potts model. Note that the standard representation is not defined for $Q = 2$, where we know from the solution of the Ising model on the Cayley tree that there is a continuous phase transition with a different exponent $\beta = 1/2$ (see above). To capture this behavior we introduce the alternating representation, $\xi = 2$, into our equations, so that the Boltzmann weight reads now: $\mu(g_v, g_{v'}) \propto 1 + \sum_{\xi=2}^{3} K_\xi \overline{\chi}_\xi(g_v^{-1} g_{v'})$, and the beliefs: $\nu(g_v) \propto [1 + \sum_{\xi=2}^{3} h_\xi^{(k)} \overline{\chi}_\xi(g_v)]^{q-1}$. Given the Abelian nature of the alternating representation, so that $\chi_2(gh) = \chi_2(g)\chi_2(h)$, and that the product of alternating × standard is another irrep, $\chi_2(g)\chi_3(g) = \chi_{2\otimes 3}(g) = \chi_4(g)$, we arrive at the following cavity field equations:

$$h_2^{(k)} = \frac{K_2}{Q!} \frac{2h_2^{(k+1)} + \left(h_3^{(k+1)}\right)^2 \left(\frac{(Q-1)^2}{Q!}\right)(1 - \delta_{Q,2})\delta_{Q,3}}{1 + \left(h_2^{(k+1)}\right)^2 \left(\frac{Q-1}{Q!}\right)^2 + \left(h_3^{(k+1)}\right)^2 \left(\frac{Q-1}{Q!}\right)^2 (1 - \delta_{Q,2})},$$

$$h_3^{(k)} = \frac{K_3}{Q!} \frac{2h_3^{(k+1)} + \left(h_3^{(k+1)}\right)^2 \left(\frac{Q-1}{Q!}\right)}{1 + \left(h_2^{(k+1)}\right)^2 \left(\frac{Q-1}{Q!}\right)^2 + \left(h_3^{(k+1)}\right)^2 \left(\frac{Q-1}{Q!}\right)^2 (1 - \delta_{Q,2})} (1 - \delta_{Q,2}). \tag{19}$$

Plugging in $Q = 2$ for the fixed point equations, we find the exponent $\beta = 1/2$, and $\beta = 1$ for $Q \neq 2$, as claimed. This is consistent with the fact that $Q = 2$ is special as the symmetry group $S_2$ is Abelian, and different from $S_Q \times S_Q$ for generic $Q$.

Note that the above results focused on the case of homogenous boundary fields, so they can only allows us to compute bulk exponents. Unfortunately, we were not able to extend these results in the presence of a DW at the boundary, as the sum $\sum_g \chi(g)\chi(g_{\text{SWAP}}g)\chi(g^{-1}h)$ (Fig. 7) cannot be simply evaluated for general replica and Renyi indices $m, n$.

## RANDOM TREE TENSOR NETWORK ENTANGLEMENT COMPUTATIONS

In this appendix, we explain the numerical methods and procedures used to verify the predictions in the main text. We generate random tree tensor networks on the rooted Cayley tree of coordination $q = 3$ and number of leaves $L = 64$, 128, and 256. For each tree, we compute the entanglement entropy (von Neumann and Renyi) for all contiguous intervals of sites $A$ with size $L_A \leq L/2$. Below we explain the methods for generating the tensors of the random tensor network and for computing the entanglement for intervals in the tree.

*Random tree tensor network ensemble*—The motivating model for this work consists of $D \times D \times D$ sized random tensors, obtained by sampling each of the $D^3$ tensor elements independently from a featureless Gaussian distribution characterized by zero mean and unit variance. However, we would like to characterize a continuous phase transition by interpolating between integer bond dimensions $D$ with a smooth parameterization. The exact choice of parameterization should not affect universal features of the critical regime $D \sim D_c$ as long as it is sufficiently smooth at the critical point.

We investigated two such parameterizations. For the first, random tensors $A$ of integer size $D \times D \times D$ were taken for each node of the tree tensor network and augmented by tensors $\lambda(\theta)$ of size $D \times D$ on each bond of the tree, as shown in Fig. 8a. The tensors $\lambda$ take the form

$$\lambda(\theta) = \begin{bmatrix} 1 & & & \\ & \ddots & & \\ & & 1 & \\ & & & \theta \end{bmatrix}.$$

By tuning $\theta$ from 0 to 1, this allows us to smoothly tune from bond dimension $D - 1$ to bond dimension $D$. This parameterization is defined piecewise between each integer bond dimension and is continuous but has kinks at the points of integer bond dimension. Our initial results for this parameterization suggest that the critical point is near the $D = 2$ point, with $D > 2$ showing log entanglement scaling and $D < 2$ showing much lower entanglement.

In order to avoid having a kink in the parameterization nearby to the critical point, we instead analyzed a second parameterization that tunes directly from $D = 1$ to $D = 3$. To achieve this, we use $3 \times 3 \times 3$ (Gaussian) random tensors $A$ at each node of the tree and bond tensors

$$\lambda(\gamma) = \begin{bmatrix} 1 & & \\ & \gamma & \\ & & \gamma^2 \end{bmatrix}.$$

The point $\gamma = 1$ corresponds to the unmodified $D = 3$ tree tensor network states, while $\gamma = 0$ corresponds to the (trival) tree network states with $D = 1$. We found a critical point in this ensemble of random trees at $\gamma = \gamma_c \approx 0.48$. Fig 8b compares the two parameterizations by showing the Renyi entanglement $S_n(\lambda)$ of the bond tensor $\lambda$ when viewed as an entangled pair state. The numerical result discussed in the main text refers to states generated from this ensemble.

*Entanglement computations*—For each random tree, we used the following procedure to compute the entanglement properties. First, each tree tensor network state is converted to an *isometric* tree tensor network state, using the redundancy of the tensor network description to modify the tensors but preserve the overall state. Each rank-3 tensor $A_{i,j}^k$ undergoes a $QR$ decomposition $A_{i,j}^k = Q_{i,j}^{k'} R_{k'}^k$ so that $Q$ is an isometry - $\left(Q_{i,j}^{k'}\right)^* Q_{i,j}^k = \delta^{k'k}$. The replacement $A_{i,j}^k \to Q_{i,j}^{k'}$ paired with the replacement $A_{k,m}^n \to R_{k'}^{k'} A_{k,m}^n$ for the tensor on the parent node preserves the overall state. Making these replacements throughout the tree, starting at the nodes closest to the leaves of the tree and progressing towards the root, converts the entire tree to isometric tensors. For such isometric tree tensor network states, the singular values of the root tensor $A_{ij}$ are the Schmidt values for the entire state corresponding to the bipartition separating the sites below bond $i$ from those below bond $j$. Each bond in the tree corresponds to a bipartition of the sites into the two sets of sites that are disconnected when erasing that bond. To compute the entanglement for these bipartitions, shift the root of the tree to these bonds using a sequence of moves as shown in Fig 9a. Each move is a contraction of the root tensor with one of its neighboring tensors in the tree followed by a $QR$-decomposition.

To compute the entanglement for other intervals $A = \{a, a+1, \ldots, b\}$ that do not correspond to cutting a single bond in the tree, we use the tree topology-changing moves shown in Fig. 9b. Each such move is performed by contracting the root tensor with both its neighbors followed by a singular value decomposition (SVD). As the bond dimensions of the tree are potentially larger after each such topology-changing move, we use truncated singular value

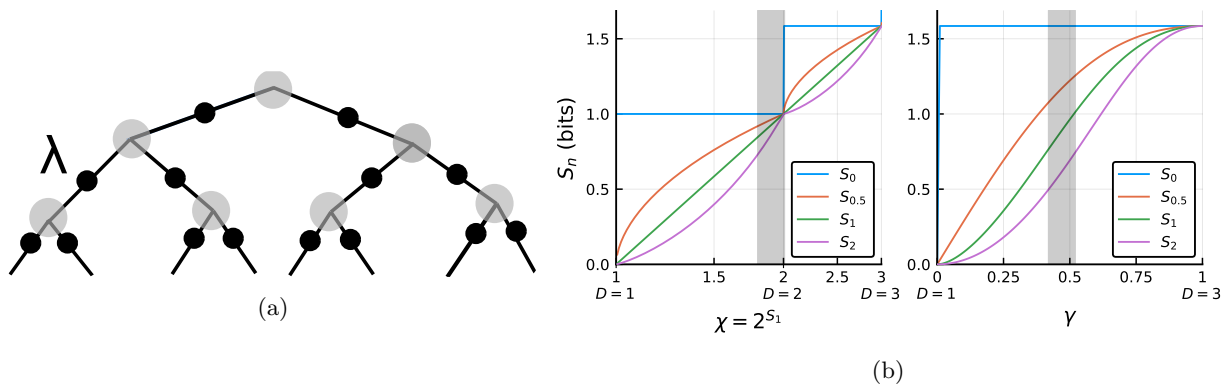

FIG. 8: (a) Bond tensors $\lambda$ are used to augment the gaussian random tensors on each node to tune continuously between integer bond dimensions. (b) Renyi entanglement of the bond tensor $\lambda$ in the two interpolations considered in the text. The first shows critical behavior near the kink at $D = 2$, the later at $\gamma \approx 0.48$ (shaded regions).

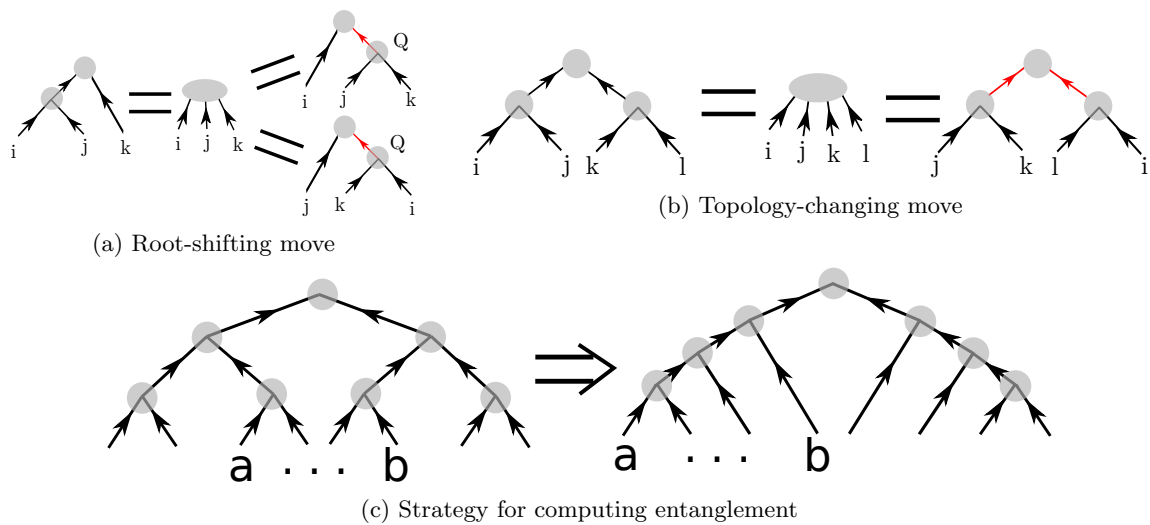

(a) Root-shifting move

(b) Topology-changing move

(c) Strategy for computing entanglement

FIG. 9: Isometric tree tensor networks consists of tensors which are isometries from incoming bonds to outgoing bonds, as denoted by the arrows. (a, b) Transformations of the network used to compute entanglement for different bonds and bipartitions corresponding to multiple bonds. (c) The entanglement for all intervals of sites starting at site $a$ can be computed using a sequence of tree topology-changing moves that morph the tree network from the initial Caylee tree to the shape on the right, which can reinterpreted as a matrix product state.

decompositions to control any unnecessary growth of the bond dimension. These truncations are done with fixed truncation error $\varepsilon = 10^{-8}$.

By using a sequence of such moves, we can create a tree in which the root tensor separates the sites of $A = \{a, a+1 \ldots b\}$ from the rest of the system. There are many valid choices of such trees — we use the tree shown in Fig. 9c, which can be interpreted as a matrix product state starting on site $a$. This tree shape allows us to also get the entanglement for intervals of all lengths starting on site $a$ using the root-shifting move in Fig 9a. For each random tree tensor network state generated, we compute the entanglement for every possible interval in the system by converting to the MPS starting on site $a$ for each site $a$. We then average the entanglement for cuts of different sizes $L_A$ over all positions of such intervals $A$ in the system that do not wrap around the boundary.



\end{document}